\documentclass[prd,nofootinbib,preprint]{revtex4}
\usepackage{latexsym} 
\usepackage{epsf}
\def\lsim{\mathrel{\rlap{\lower4pt\hbox{\hskip1pt$\sim$}}
    \raise1pt\hbox{$<$}}}      %less than or approx. symbol
\def\gsim{\mathrel{\rlap{\lower4pt\hbox{\hskip1pt$\sim$}}
    \raise1pt\hbox{$>$}}}      %greater than or approx. symbol
\begin{document}
%\preprint{\hbox{YITP-SB-54-03}}

\title{Spontaneous CP violation in the 3-3-1 model with right-handed neutrinos}

\author{A. Doff}
\affiliation{Instituto de  F\'{\i}sica Te\'orica, Universidade
Estadual Paulista,\\ Rua Pamplona 145, 01405-000, S\~ao Paulo-SP, Brazil.}
\author{C. A. de S. Pires}
\affiliation{Departamento de F\'{\i}sica, Universidade Federal da
Para\'{\i}ba, Caixa Postal 5008, 58051-970, Jo\~ao Pessoa - PB,
Brazil.}
\author{P. S. Rodrigues da Silva}
\affiliation{Departamento de F\'{\i}sica, Universidade Federal da
Para\'{\i}ba, Caixa Postal 5008, 58051-970, Jo\~ao Pessoa - PB,
Brazil.}
%---------------------------------------------------------------------

\begin{abstract}
\vspace*{0.5cm}
We implement the mechanism of spontaneous CP violation in the 3-3-1 model with right-handed neutrinos and recognize their sources  of CP violation. Our main result is that the  mechanism works already in the minimal version of the model and new sources of CP violation emerges as an effect of new physics at energies higher than the electroweak scale.
%\noindent PACS numbers: 12.60.Cn; 11.30.Er.
\end{abstract}

\maketitle

%---------------------------------------------------------------------
\section{Introduction}
\label{sec1}
There are two possibilities of implementation of CP violation in gauge models. The most popular one is the  explicit CP violation mechanism~\cite{KM}. This is the case of  CP violation  in the quark sector of the standard model (SM) of the electroweak interactions. Basically it is assumed that the Yukawa parameters of the model are  complex, thus providing a complex mass matrix for the quarks whose diagonalization requires unitary transformation. As a consequence, considering three generations of quarks, one arbitrary phase will remain in the CKM mixing matrix~\cite{KM,cabibbo}. As such a mixing appears only in the quark sector that is the only place where CP violation can arise in SM~\cite{coment1}.

Another way to obtain CP violation is through the spontaneous breaking of the gauge symmetry. In gauge models the breaking of the gauge symmetry is achieved through a non-zero vacuum expectation value (VEV) of some neutral scalars. In the general case such VEV should be complex. In this case, even though we departure from a real Lagrangian, when the neutral scalars of the model develop VEV's, phases will be introduced into the Lagrangian. Some of these phases can remain in the model depending on the constraints of the scalar potential at its minimum and whether the fields redefinition do not eat them. If at least one phase survives, we have spontaneous CP violation (SCPV)~\cite{lee}.

Although  more appealing than the explicit case, it is not straightforward to implement the SCPV mechanism in gauge models. Usually  the phases  that accompany the real part of the VEV큦 can be easily  eliminated by an adequate   gauge transformation. For example,  in the standard model the $SU(2)_L \otimes U(1)_Y$ gauge symmetry allows us to eliminate the phase of the standard VEV. The minimal extension of the standard model that accommodates SCPV does it through the addition of a second Higgs doublet in the context  of the most general potential~\cite{CPsm}.

The procedure to implement the SCPV in any model is quite simple. Consider all neutral scalars that can  develop complex VEV in the respective model. Obtain the constraint equations on the phases which come from the minimum condition of the scalar potential. Check how many phases the gauge transformation can eliminate. Finally, wherever it is possible, redefine the fields as to absorb the remaining phases. If, in the end, at least one phase survives, then the SCPV mechanism works in the model considered.

In this work we implement the SCPV mechanism in the $SU(3)_C \otimes SU(3)_L \otimes U(1)_N$ (3-3-1) gauge model for the electroweak and strong interactions that posses right-handed neutrinos composing the fundamental representation~\cite{themodel}. The SCPV was already considered in other versions of 3-3-1 models~\cite{dumm,vicente}. In general all the 3-3-1 models require at least three scalar triplets in order to generate the correct particle masses of the model. Considering that each scalar triplet involves at least one neutral scalar, then we are going to have at least three phases when no complex coupling is assumed. Nevertheless, the $SU(3)_L \otimes U(1)_N$ symmetry allows us to eliminate two of these phases and, certainly, the minimum condition of the scalar potential of the model will provide an additional constraint over the phases.  Thus, it is expected that, in the case of three triplets and three phases, the gauge symmetry and the minimum constraint lead to a trivial solution, i.e., that these three phases can be washed out of the model. It was shown in Ref.~\cite{dumm} that this kind of 3-3-1 model requires three triplets and one scalar sextet in order to SCPV be implemented. 

The minimal 3-3-1 model with right-handed neutrinos also requires only three scalar triplets to generate the needed masses. What motivates our investigation of the SCPV in this model is that its scalar sector differs from the other models in a subtle aspect. Two of its three scalar triplets posses each one  two neutral scalars. Thus, in this model, we dispose from the beginning of a total of five neutral scalars. When we assume that all of them develop complex VEV's, five phases appear in the model. We will show that in this scenario one phase still survives the  gauge transformation, the constraints for the minimum of the potential and the fields redefinition, allowing for SCPV mechanism with only three scalar triplets.

This paper is organized as follows. In Sec. II we specify the model. We then implement SCPV in this 3-3-1 model in Sec. III. In Secs. IV and V we point out the sources of CP violation. We finish with our conclusions in Sec. VI.
\section{the model}

In the 3-3-1 model with right-handed neutrinos the leptons come in triplet and in singlet representations,
\begin{eqnarray}
f_{aL} = \left (
\begin{array}{c}
\nu_{aL} \\
e_{aL} \\
\nu^C_{aL}
\end{array}
\right )\sim(1\,,\,3\,,\,-1/3)\,,\,\,\,e_{aR}\,\sim(1,1,-1),
 \end{eqnarray}
while in the quark sector, one generation comes in the triplet and the other two compose
an anti-triplet representation with the following content,
\begin{eqnarray}
&&Q_{iL} = \left (
\begin{array}{c}
d_{i} \\
-u_{i} \\
d^{\prime}_{i}
\end{array}
\right )_L\sim(3\,,\,\bar{3}\,,\,0)\,,u_{iR}\,\sim(3,1,2/3),\,\,\,\nonumber \\
&&\,\,d_{iR}\,\sim(3,1,-1/3)\,,\,\,\,\, d^{\prime}_{iR}\,\sim(3,1,-1/3),\nonumber \\
&&Q_{3L} = \left (
\begin{array}{c}
u_{3} \\
d_{3} \\
u^{\prime}_{3}
\end{array}
\right )_L\sim(3\,,\,3\,,\,1/3),u_{3R}\,\sim(3,1,2/3),\nonumber \\
&&\,\,d_{3R}\,\sim(3,1,-1/3)\,,\,u^{\prime}_{3R}\,\sim(3,1,2/3)
\label{quarks} 
\end{eqnarray}
where $a = 1,\,2,\, 3$ refers to the three generations and $i=1,2$. The primed quarks
are the new quarks but with the usual electric charges.

In order to generate the correct particle masses,
the model requires only  three scalar triplets, namely,
\begin{eqnarray}
\eta = \left (
\begin{array}{c}
\eta^0 \\
\eta^- \\
\eta^{\prime 0}
\end{array}
\right ),\,\rho = \left (
\begin{array}{c}
\rho^+ \\
\rho^0 \\
\rho^{\prime +}
\end{array}
\right ) ,\, \chi = \left (
\begin{array}{c}
\chi^0 \\
\chi^{-} \\
\chi^{\prime 0}
\end{array}
\right ) , \label{scalarcont} 
\end{eqnarray}
with $\eta$ and $\chi$ both transforming as $(1\,,\,3\,,\,-1/3)$
and $\rho$ transforming as $(1\,,\,3\,,\,2/3)$.

We assume the following set of discrete symmetries over the full Lagrangian in order to have a minimal model,
\begin{eqnarray}
	\left( \chi\,,\,\eta\,,\,\rho\,,e_{aR}\,,\, u_{aR}\,,\,u^{\prime}_{3R}\,,\,d_{aR}\,,\,d^{\prime}_{iR}\right) \rightarrow -\left( \chi\,,\,\eta\,,\,\rho\,,e_{aR}\,,\, u_{aR}\,,\,u^{\prime}_{3R}\,,\,d_{aR}\,,\,d^{\prime}_{iR}\right)
	\label{discretesymmetryI}
\end{eqnarray}
This set of discrete symmetries helps in avoiding the undesirable Dirac mass terms for the neutrinos and  allows a minimal and realistic CP conserving potential~\cite{pal},
\begin{eqnarray} V(\eta,\rho,\chi)&=&\mu_\chi^2 \chi^2 +\mu_\eta^2\eta^2
+\mu_\rho^2\rho^2+\lambda_1\chi^4 +\lambda_2\eta^4
+\lambda_3\rho^4+ \nonumber \\
&&\lambda_4(\chi^{\dagger}\chi)(\eta^{\dagger}\eta)
+\lambda_5(\chi^{\dagger}\chi)(\rho^{\dagger}\rho)+\lambda_6
(\eta^{\dagger}\eta)(\rho^{\dagger}\rho)+ \nonumber \\
&&\lambda_7(\chi^{\dagger}\eta)(\eta^{\dagger}\chi)
+\lambda_8(\chi^{\dagger}\rho)(\rho^{\dagger}\chi)+\lambda_9
(\eta^{\dagger}\rho)(\rho^{\dagger}\eta)+ \nonumber \\
&&\frac{f}{\sqrt{2}}\epsilon^{ijk}\eta_i \rho_j \chi_k +\mbox{H.c}.
\label{potential}
\end{eqnarray}
The Yukawa sector of the model is given by the  following CP conserving interactions,
\begin{eqnarray}
	{\cal L}^Y &=&f_{ij}\left( \bar Q_{iL}\chi^* d^{\prime}_{jR}+\bar Q_{iL}\eta^* d^{\prime}_{jR} \right)+f_{33}\left( \bar Q_{3L}\chi u^{\prime}_{3R}+\bar Q_{3L}\eta u^{\prime}_{3R} \right)+ g_{ia}\left( \bar Q_{iL}\eta^* d_{aR}+\bar Q_{iL}\chi^* d_{aR} \right) \nonumber \\
	&&+h_{3a}\left( \bar Q_{3L}\eta u_{aR}+\bar Q_{3L}\chi u_{aR} \right)+g_{3a}\bar Q_{3L}\rho d_{aR}+h_{ia}\bar Q_{iL}\rho^* u_{aR}+ G_{aa}\bar f_{aL} \rho e_{aR}+\mbox{H.c}.
\label{yukawainteractions}
\end{eqnarray}
With this set of Yukawa interactions all fermions, except the neutrinos, gain masses. In this model neutrinos gain masses through effective dimension-five operators as shown in Ref.~\cite{sterile}.
\section{Spontaneous CP violation}
Let us assume that  all neutral scalars composing the triplets, $\rho$ , $\eta$ and $\chi$, develop complex VEV's,
{\small
\begin{eqnarray}
\eta = \left (
\begin{array}{c}
\frac{e^{i\theta_\eta}}{\sqrt{2}}\left( v_\eta+R_\eta +iI_\eta \right) \\
\eta^- \\
\frac{e^{i\theta_{\eta^{\prime}}}}{\sqrt{2}}\left( v_{\eta^{\prime}}+R_{\eta^{\prime}} +iI_{\eta^{\prime}} \right)
\end{array}
\right ),\,\rho = \left (
\begin{array}{c}
\rho^+ \\
 \frac{e^{i\theta_\rho}}{\sqrt{2}}\left( v_\rho+R_\rho +iI_\rho \right)\\
\rho^{\prime +}
\end{array}
\right ) ,\, \chi = \left (
\begin{array}{c}
\frac{e^{i\theta_\chi}}{\sqrt{2}}\left(v_\chi +R_\chi + iI_\chi \right) \\
\chi^- \\
\frac{e^{i\theta_{\chi^{\prime}}}}{\sqrt{2}}\left(v_{\chi^{\prime}} +R_{\chi^{\prime}} + iI_{\chi^{\prime}} \right)
\end{array}
\right ). \label{VEV} 
\end{eqnarray}
}
The minimum condition for the potential in Eq.(\ref{potential}), when the scalars develop five non-trivial VEV's,  translates into a set of  constraints on the VEV's and phases of the model.  On substituting Eq.(\ref{VEV}) at the  potential in Eq.~(\ref{potential}), we obtain the following set of constraints on the phases,
\begin{eqnarray}
	\theta_\eta + \theta_\rho + \theta_{\chi^{\prime}}=0 \,\,\,\,\,,\,\,\,\,\,	\theta_{\eta^{\prime}} + \theta_\rho + \theta_{\chi}=0,
	\label{phaseconstraint}
\end{eqnarray}
and the following  ones on the VEVs,
\begin{eqnarray} 
&&\mu^2_\chi +\lambda_1 \left(v^2_{\chi^{\prime}}+v^2_\chi \right) +
\frac{\lambda_4}{2}\left( v^2_\eta + v^2_{\eta^{\prime}} \right) +
\frac{\lambda_5}{2}v^2_\rho+\frac{\lambda_7}{2}\left(
v_{\eta^{\prime}}v_{\chi^{\prime}} + v_\eta v_\chi \right)\frac{v_\eta}{v_\chi}-\frac{f}{2}\frac{v_{\eta^{\prime}} v_\rho}
{ v_{\chi}}=0,\nonumber \\
&&\mu^2_\chi +\lambda_1 \left(v^2_{\chi^{\prime}}+v^2_\chi \right) +
\frac{\lambda_4}{2}\left( v^2_\eta + v^2_{\eta^{\prime}} \right) +
\frac{\lambda_5}{2}v^2_\rho+\frac{\lambda_7}{2}\left(
v_{\eta^{\prime}}v_{\chi^{\prime}} + v_\eta v_\chi \right)\frac{v_{\eta^{\prime}}}{v_{\chi^{\prime}}}+\frac{f}{2}\frac{v_{\eta} v_\rho}
{ v_{\chi^{\prime}}}=0,
\nonumber \\
&&\mu^2_\eta +\lambda_2 \left( v^2_{\eta^{\prime}} + v^2_\eta \right) +
\frac{\lambda_4}{2}\left( v^2_{\chi^{\prime}}+v^2_\chi \right)
 +\frac{\lambda_6}{2}v^2_\rho +\frac{\lambda_7}{2}\left( v_\chi v_\eta + v_{\chi^{\prime}}v_{\eta^{\prime}} \right)\frac{v_\chi}{v_\eta}
+\frac{f}{2}\frac{v_\rho v_{\chi^{\prime}}}{v_\eta} =0,\nonumber \\&&
\mu^2_\eta +\lambda_2\left( v^2_{\eta^{\prime}} + v^2_\eta \right) +
\frac{\lambda_4}{2}\left( v^2_{\chi^{\prime}}+v^2_\chi \right)
 +\frac{\lambda_6}{2}v^2_\rho +\frac{\lambda_7}{2}\left( v_\chi v_\eta + v_{\chi^{\prime}}v_{\eta^{\prime}} \right)\frac{v_{\chi^{\prime}}}{v_{\eta^{\prime}}}
-\frac{f}{2}\frac{v_\rho v_{\chi}}{v_{\eta^{\prime}}} =0,\nonumber \\&&
\mu^2_\rho +\lambda_3 v^2_\rho + \frac{\lambda_5}{2}\left(
v^2_{\chi^{\prime}}+v^2_\chi \right) +\frac{\lambda_6}{2}( v^2_{\eta^{\prime}}+
v^2_\eta)+\frac{f}{2}\frac{v_\eta v_{\chi^{\prime}}}{v_\rho}-\frac{f}{2}\frac{v_{\eta^{\prime}} v_\chi }{v_\rho} =0.
	\label{VEVconstraints}
\end{eqnarray}
The origin of the constraints on the phases is the non-hermitian $f$-term  in the potential in Eq.~(\ref{potential}). Observe that we have only two constraints on the  five phases.  This is not sufficient for us to conclude whether we have a non-trivial solution or not because the $SU(3)_L \times U(1)_N$ symmetry allows us to take two of these phases  as zero. This, together with the constraint in Eq.(\ref{phaseconstraint}), allow us to get rid of four phases. As we have five phases, in the end of the day  one phase survives, and we could possibly end up with SCPV if, after fields redefinition, such a phase cannot be eliminated.

Note that it is essential for the SCPV that the model develops five phases. We have to emphasize that the model works perfectly well when only three of the neutral scalars, $\eta^0$, $\rho^0$  and $\chi^{\prime 0}$, develop complex VEV큦. In this case the only constraint that survives is the first one in Eq.(\ref{phaseconstraint}). However this constraint together with the gauge freedom are sufficient to eliminate the three phases. This is also the case with four VEV큦. Thus SCPV in the 3-3-1 model with right-handed neutrinos necessarily requires that all neutral scalars of the three triplets of the model develop VEV큦.

It is important to stress that in the 3-3-1 model presented in the second article of Ref.~\cite{vicente} the equivalent trilinear term explicitly violates CP. As we will see below, there is no need of any kind of explicit CP violation here, so that our procedure might yield a genuine SCPV mechanism, without recurring to a CP violation coupling or any extra scalar fields neither a huge potential.

Once we are sure that at least one phase remains after the procedure above described, the next step is to obtain the sources of CP violation in the model. This is done by opening up all interaction terms in the model and caring about the fields rotation we are allowed to make, in order to render mass matrices real. If after all possible fields redefinition a phase still survives, we promptly identify the terms in the Lagrangian which violate CP. In view of this it is adequate, at this point of the work, to choose which phases we are going to take as zero. Without any particular reason  we can take
\begin{eqnarray}
 \theta_\rho=\theta_\eta=0\,.
 \label{choice}
 \end{eqnarray}
With this, the constraints in Eq.(\ref{phaseconstraint}) automatically imposes 
\begin{eqnarray}
\theta_{\chi^{\prime}}=0\,\,\,,\,\,\, \mbox{and}\,\,\,,\,\,\,\theta_\chi=-\theta_{\eta^{\prime}}.
\label{phaseimplication}
\end{eqnarray}

Next we are going to explore the consequences of this framework.
%
%%%%%%%%%%%%%%%%%%%%%%%%%%%%%%%%%%%%%%%%%%%%%%%%%%%%%%%%%%%%%%%%%%%%%%%%%%%%%%%%%%%%%%%%%%%%%%%%%%%%%%%%%%%%%%%%%%%%%%%%%%%%%%%%%%%%%%
\section{Quark sector}
We first  concentrate on the  up-type quarks. Expanding the neutral scalars about their respective VEV's, given by Eq.~(\ref{VEV}), and using the results in Eqs.~(\ref{choice})  and (\ref{phaseimplication}), we obtain the following mass matrix for the up-type quarks in the basis $\left ( u_1\,\, , \,\, u_2 \,\, , \,\, u_3 \,\, , \,\, u^{\prime}_3 \right)$
\begin{eqnarray}\begin{footnotesize}M_u=\frac{1}{\sqrt{2}}\left(\begin{array}{ccccc} 
 -h_{11}v_\rho & -h_{12}v_\rho  & -h_{13}v_\rho & 0\\
 -h_{21}v_\rho & -h_{22}v_\rho & -h_{23}v_\rho & 0 \\
 h_{31}(v_\eta +e^{i\theta_\chi}v_\chi) & h_{32}(v_\eta +e^{i\theta_\chi}v_\chi) & h_{33}(v_\eta +e^{i\theta_\chi}v_\chi) & f_{33}(v_\eta +e^{i\theta_\chi}v_\chi) \\
 h_{31}(v_{\chi^{\prime}}+e^{-i \theta_\chi}v_{\eta^{\prime}}) & h_{32}(v_{\chi^{\prime}}+e^{-i \theta_\chi}v_{\eta^{\prime}})  & h_{33}(v_{\chi^{\prime}}+e^{-i \theta_\chi}v_{\eta^{\prime}})  & f_{33}(v_{\chi^{\prime}}+e^{-i \theta_\chi}v_{\eta^{\prime}}) 
\end{array}
\right). \label{massmatrxU} \end{footnotesize}\end{eqnarray}

It is not possible to redefine any quark field in order to turn this matrix real. Thus in being  a complex mass matrix, it can be brought to a diagonal form by an unitary transformation, $V^u_L M_u V^{uT}_R = m_u = \mbox{diag}\left( m_u\,,\,m_c\,,\,m_t\,,\,m_T \right)$, which connects the flavor eigenstates with the  mass eigenstates
\begin{eqnarray}
\left (
\begin{array}{c}
u_{1L} \\
u_{2L} \\
u_{3L} \\
u^{\prime}_{3L}
\end{array}
\right )=
V^u_L\left (
\begin{array}{c}
u_L\\
c_L \\
t_L \\
T_L
\end{array}
\right )\,\,\,\,\,\, , \,\,\,\,\,\,\left (
\begin{array}{c}
u_{1R} \\
u_{2R} \\
u_{3R} \\
\hat{u}^{\prime}_{3R}
\end{array}
\right )=
V^u_R\left (
\begin{array}{c}
u_R\\
c_R \\
t_R \\
T_R
\end{array}
\right ).
\label{unitaryUtransformation}
 \end{eqnarray}
We represent these new bases (mass eigenstates) as $U_L= \left( u_L\,,\,c_L\,,\,t_L\,,\,T_L \right)^T$  and $U_R=\left( u_R\,,\,c_R\,,\,t_R\,,\,T_R \right)^T$.

In regard to the down-type quarks, we obtain the following mass matrix  in the basis $\left( d_1 \,\, , \,\, d_2 \,\, , \,\, d_3 \,\, , \,\, d^{\prime}_1 \,\, ,\,\, d^{\prime}_2  \right)$,
\begin{eqnarray}\begin{scriptsize}M_d=\frac{1}{\sqrt{2}}\left(\begin{array}{ccccc}
 g_{11}(v_\eta + e^{-i\theta_\chi}v_\chi) & g_{12}(v_\eta + e^{-i\theta_\chi}v_\chi)   & g_{13}(v_\eta + e^{-i\theta_\chi}v_\chi)  & f_{11}(v_\eta + e^{-i\theta_\chi}v_\chi) & f_{12}(v_\eta + e^{-i\theta_\chi}v_\chi)  \\
 g_{21}(v_\eta + e^{-i\theta_\chi}v_\chi)& g_{22}(v_\eta + e^{-i\theta_\chi}v_\chi) & g_{23}(v_\eta + e^{-i\theta_\chi}v_\chi) & f_{21}(v_\eta + e^{-i\theta_\chi}v_\chi) & f_{22}(v_\eta + e^{-i\theta_\chi}v_\chi) \\
  g_{31}v_\rho & g_{32} v_\rho & g_{33}v_\rho & 0 & 0\\
 g_{11}(v_{\chi^{\prime}}+e^{i\theta_\chi}v_{\eta^{\prime}}) & g_{12}(v_{\chi^{\prime}}+e^{i\theta_\chi}v_{\eta^{\prime}}) &g_{13}(v_{\chi^{\prime}}+e^{i\theta_\chi}v_{\eta^{\prime}}) &f_{11}(v_{\chi^{\prime}}+e^{i\theta_\chi}v_{\eta^{\prime}})&f_{12}(v_{\chi^{\prime}}+e^{i\theta_\chi}v_{\eta^{\prime}})  \\
 g_{21}(v_{\chi^{\prime}}+e^{i\theta_\chi}v_{\eta^{\prime}}) &g_{22}(v_{\chi^{\prime}}+e^{i\theta_\chi}v_{\eta^{\prime}}) &g_{23}(v_{\chi^{\prime}}+e^{i\theta_\chi}v_{\eta^{\prime}}) &f_{21}(v_{\chi^{\prime}}+e^{i\theta_\chi}v_{\eta^{\prime}}) &f_{22}(v_{\chi^{\prime}}+e^{i\theta_\chi}v_{\eta^{\prime}}) 
\end{array}
\right). \label{massmatrixd} \end{scriptsize}\end{eqnarray}

Again, this matrix cannot be made  real by redefinition of the quark fields, thus it will be diagonalized by an unitary transformation $V^d_L M_d V^{dT}_R = m_u = \mbox{diag}\left( m_d\,,\,m_s\,,\,m_b\,,\,m_D\,,\,m_S\right)$, where
\begin{eqnarray}
\left (
\begin{array}{c}
d_{1L} \\
d_{2L} \\
d_{3L} \\
d^{\prime}_{1L} \\
d^{\prime}_{2L}
\end{array}
\right )=
V^d_L\left (
\begin{array}{c}
d_L\\
s_L \\
b_L \\
B_L \\
S_L
\end{array}
\right )\,\,\,\,\,\, , \left (
\begin{array}{c}
d_{1R} \\
d_{2R} \\
d_{3R} \\
d^{\prime}_{1R}\\
d^{\prime}_{2R}
\end{array}
\right )=
V^d_R\left (
\begin{array}{c}
d_R\\
s_R \\
b_R \\
B_R \\
S_R
\end{array}
\right ),
\label{unitaryDtransformation}
 \end{eqnarray}
where the basis for the mass eigenstates is represented by $D_L= \left( d_L\,,\,s_L\,,\,b_L\,,\,B_L\,,\,S_L \right)^T$  and $D_R=\left( d_R\,,\,s_R\,,\,b_R\,,\,B_R\,,\,S_R \right)^T$. 
 
\subsection{Quark-neutral scalars couplings} 
The neutral current interactions among the up-type quarks and the neutral scalars of the model are given by,
\begin{eqnarray}
{\cal L}^{u-u-neutral}_Y	=\bar U_{\beta L}A^{\prime \prime}_{\beta \alpha k}U_{\alpha R} H_k +i\bar U_{\beta L}B^{\prime \prime}_{\beta \alpha k}U_{\alpha R} G_k +\mbox{H.c.},
\label{neutralcurrentU}
\end{eqnarray}
where
\begin{eqnarray}
A^{\prime \prime}_{\beta \alpha k}=&&	\left( \frac{f_{33}}{\sqrt{2}}V^{u \dagger}_{3 \beta L}V^u_{4\alpha R} +\frac{h_{3a}}{\sqrt{2}}V^{u \dagger}_{3 \beta L}V^u_{a \alpha R} \right) \left( e^{i\theta_\chi}{\cal O}^R_{4 k}+e^{-i\theta_\chi}{\cal O}^R_{5k}+{\cal O}^R_{2k}+{\cal O}^R_{1k}\right)H_k \nonumber \\
&&-\frac{h_{ia}}{\sqrt{2}}V^{u \dagger}_{i \beta L}V^u_{a \alpha R}{\cal O}^R_{3k}H_k,
\nonumber 
\end{eqnarray}
\begin{eqnarray}
B^{\prime \prime}_{\beta \alpha k}=&&\left( \frac{f_{33}}{\sqrt{2}}V^{u \dagger}_{3 \beta L}V^u_{4\alpha R} +\frac{h_{3a}}{\sqrt{2}}V^{u \dagger}_{3 \beta L}V^u_{a \alpha R} \right) \left( e^{i\theta_\chi}{\cal O}^I_{4 k}+e^{-i\theta_\chi}{\cal O}^I_{5k}+{\cal O}^I_{2k}+{\cal O}^I_{1k}\right)G_k \nonumber \\
&&-\frac{h_{ia}}{\sqrt{2}}V^{u \dagger}_{i \beta L}V^u_{a \alpha R}{\cal O}^I_{3k}G_k .
\label{Aneutralreal}
\end{eqnarray}
where $\beta \,,\, \alpha\,,\,\,k=1,2,3,4$ and $H_k$, $G_k$  are the neutral scalars mass eigenstates and the factors ${\cal O}^R$ and ${\cal O}^I$ are elements of the diagonalization matrices for the scalars (see the appendix).

Notice that in Eqs.~(\ref{Aneutralreal}), there are phases multiplying some of the real factors (those related to elements of orthogonal matrices ${\cal O}$'s) but not all. Thus, even if the products of the diagonalization matrix elements present in these equations were not complex, this explicit phase guarantees that $A^{\prime \prime}_{\beta \alpha k}$ and $B^{\prime \prime}_{\beta \alpha k}$ carry terms with distinct phases.
Then, the neutral interactions in Eq.~(\ref{neutralcurrentU}) involve flavor changing and CP violation as is usual in models of SCPV.

The interactions of the  physical down-type quarks  with the neutral scalars are given by
\begin{eqnarray}
{\cal L}^{d-d-neutral}_Y=\bar D_{\gamma L} A^{\prime}_{\gamma \sigma k}D_{\sigma R}H_k -i \bar D_{\gamma L}B^{\prime}_{\gamma \sigma k}D_{\sigma R}G_k +\mbox{H.c},
\label{neutralDint}
\end{eqnarray}
where,
\begin{eqnarray}
	A^{\prime}_{\gamma \sigma k}=&&\left( \frac{f_{ip}}{\sqrt{2}}V^{d \dagger}_{i \gamma L}V^d_{p \sigma R}+ \frac{g_{ia}}{\sqrt{2}}V^{d \dagger}_{i \gamma L}V^d_{a \sigma R} \right)\left( e^{-i\theta_\chi}{\cal O}^H_{4k}+{\cal O}^H_{2k} \right)+\nonumber \\
	&&\left( \frac{f_{pq}}{\sqrt{2}}V^{d \dagger}_{p \gamma L}V^d_{q \sigma R} +\frac{g_{pa}}{\sqrt{2}}V^{d \dagger}_{p \gamma L}V^d_{a\sigma R} \right)\left( e^{i\theta_\chi}{\cal O}^H_{5k}+{\cal O}^H_{1k} \right)+\frac{g_{3a}}{\sqrt{2}}V^{d \dagger}_{3 \gamma L}V^d_{a \sigma R}{\cal O}^H_{3k},
	\nonumber
\end{eqnarray}
\begin{eqnarray}
B^{\prime}_{\gamma \sigma k}=&&\left( \frac{f_{ip}}{\sqrt{2}}V^{d \dagger}_{i \gamma L}V^d_{p \sigma R}+ \frac{g_{ia}}{\sqrt{2}}V^{d \dagger}_{i \gamma L}V^d_{a \sigma R} \right)\left( e^{-i\theta_\chi}{\cal O}^I_{4k}+{\cal O}^I_{2k} \right)+\nonumber \\
	&&\left( \frac{f_{pq}}{\sqrt{2}}V^{d \dagger}_{p \gamma L}V^d_{q \sigma R} +\frac{g_{pa}}{\sqrt{2}}V^{d \dagger}_{p \gamma L}V^d_{a\sigma R} \right)\left( e^{i\theta_\chi}{\cal O}^I_{5k}+{\cal O}^I_{1k} \right)+\frac{g_{3a}}{\sqrt{2}}V^{d \dagger}_{3 \gamma L}V^d_{a \sigma R}{\cal O}^I_{3k}.
	\label{coefneutralD}	
\end{eqnarray}
where $\gamma,\sigma=1,2,3,4,5$, $p,q=4,5$, $i=1,2$  and $a=1,2,3$. As before, similarly to the up-quarks interactions,  a phase survives here and we have CP violation coming from  interactions presenting flavor changing neutral currents.

\subsection{Quark-charged scalars couplings}

Let us now present the Yukawa interactions among quarks and charged scalars. Making use of the redefinition of the  charged scalars  $\eta^+$  and $\rho^+$ given in Eq.~(\ref{redchargedscalar}), the Yukawa interactions among physical quarks and charged scalars take the simple form
\begin{eqnarray}
		{\cal L}^{u-d-h^+}_Y&=&\bar{U}_\beta A_{\beta \gamma k}\frac{1+\gamma_5}{2}D_\gamma h^+_k + \bar{U}_\beta B_{\beta \gamma k}\frac{1-\gamma_5}{2}D_\gamma h^+_k+\mbox{H.c},
	\label{CPchargedscalarquarks}	
\end{eqnarray}
where
\begin{eqnarray}
	A_{\beta \gamma r}&=&-f_{ip}V^{u \dagger}_{i \beta L}V^d_{p \gamma R}
{\cal O}^h_{1k}-g_{ia}V^{u\dagger}_{i\beta L}V^d_{a \gamma R}{\cal O}^h_{1k}
-e^{i\theta_\chi}f_{ip}V^{u \dagger}_{i \beta L}V^d_{p \gamma R}{\cal O}^h_{2k}
\nonumber \\
&&-e^{i\theta_\chi}g_{ia}V^{u\dagger}_{i\beta L}V^d_{a \gamma R}{\cal O}^h_{2k}
+e^{i\theta_\chi}g_{3a}V^{u\dagger}_{3 \beta L}V^d_{a \gamma R}{\cal O}^h_{3k}+g_{3a}V^{u\dagger}_{4 \beta L}
V^d_{a \gamma R}{\cal O}^h_{4k},
\nonumber \\
B_{\beta \gamma r}&=&f_{33} V^{u \dagger}_{4 \beta R}V^d_{3 \gamma L}
{\cal O}^h_{1k}+h_{3a}V^{u\dagger}_{a\beta R}V^d_{3 \gamma L}{\cal O}^h_{1k}
+e^{i\theta_\chi}f_{33} V^{u \dagger}_{4 \beta R}V^d_{3 \gamma L}{\cal O}^h_{2k}
\nonumber \\
&&+e^{i\theta_\chi}h_{3a}V^{u\dagger}_{a\beta R}V^d_{3 \gamma L}{\cal O}^h_{2k}
+e^{i\theta_\chi}h_{ia}V^{u\dagger}_{a \beta R}V^d_{i \gamma L}{\cal O}^h_{3k}
+h_{pa}V^{u\dagger}_{a \beta R}V^d_{p \gamma L}{\cal O}^h_{4k}.
\nonumber \\
\label{AB}
\end{eqnarray} 
Looking at the several terms in these equations, we reach the same conclusion as before, the surviving phase establishes that we have CP violation involving quarks and charged scalars too. 

\subsection{Quark-gauge boson couplings}

Finally, we consider the gauge boson interactions with quarks. The 3-3-1  model disposes of nine gauge bosons. Four of them are charged, $W^{\pm}$ and $V^{\pm}$, plus a pair of neutral non-hermitian gauge bosons, $U^0$and $U^{0 \dagger}$, and  three neutral hermitian ones, $Z^0$, $Z^{\prime 0}$  and the photon, $\gamma$.

In view of the quark mixing given in  Eq.~(\ref{unitaryUtransformation})  and Eq.~(\ref{unitaryDtransformation}), the interactions involving the  standard charged gauge bosons, $W^{\pm}$, and the quarks  are given by the  terms
\begin{eqnarray}
{\cal L}^{cc}_{W}=-\frac{g}{\sqrt{2}}\bar u_L V_{CKM}\gamma^\mu d_L W^+_\mu -\frac{g}{\sqrt{2}}\bar T_L\left( V^{u\dagger}_{34L}V^d_{3pL} + V^{u\dagger}_{i4L}V^d_{ipL} \right) \gamma^\mu D_{pL}W^+_\mu +\mbox{H.c},
	\label{Wud}
\end{eqnarray}
where $u=\left( u\,,\,c\,,\,t \right)^T$ , $d=\left( d\,,\,s\,,\,b \right)^T$, $i=1,2$, $p=4,5$, $\beta=1,2,3,4$ and  $V_{CKM}=\left( V^{u\dagger}_{i\beta L}V^d_{i\beta L}+V^{u\dagger}_{3\beta L}V^d_{3\beta L} \right)$  can be arranged to mimic the standard CKM mixing matrix which can be parametrized by three angles and one phase. Then, as usual, the first interaction term in Eq.~({\ref{Wud}) presents CP violation because $V_{CKM}$ disposes of a phase. This term  recovers all the standard CP violation implications as, for instance, the standard model contribution to the $\epsilon$ parameter in the $K^0-\bar K^0$ system. The last interaction term involving $W^{\pm}$ and the new quarks $B$, $S$  and $T$ arises due to the mixing  among the standard quarks and the new quarks. 

The interactions involving  the $V^{\pm}$ gauge bosons and quarks are given by
\begin{eqnarray}
{\cal L}^{cc}_V=-\frac{g}{\sqrt{2}}\left( \bar{U}_{\beta L}V^{u\dagger}_{4 \beta L}V^d_{3 \gamma L}\gamma^\mu D_{\gamma L} + \bar{U}_{\beta L}V^{u\dagger}_{i \beta L}V^d_{p \gamma L}\gamma^\mu D_{\gamma L}\right) V^+_\mu.
\label{Vquarks}	
\end{eqnarray}
However when the neutral scalars $\chi^0$  and $\eta^{\prime 0}$ develop their respective VEV큦 a mixing among $W^{+}$  and $V^+$ arises
\begin{eqnarray}
\left( W^+_\mu\,,\,V^+_\mu \right) \frac{g^2}{4}\left( \begin {array}{cc} 
v^2_\rho + v^2_\chi + v^2_\eta & e^{-i\theta_\chi}(v_\chi v_{\chi^{\prime}} +v_\eta v_{\eta^{\prime}})\\
e^{i\theta_\chi}(v_\chi v_{\chi^{\prime}} +v_\eta v_{\eta^{\prime}}) & v^2_\rho +v^2_{\eta^{\prime}}+v^2_{\chi^{\prime}}
\end {array} \right)	\left (
\begin{array}{c}
W^-_\mu \\
V^-_\mu
\end{array}
\right ).
\label{chargedgaugebosonmassmatrix}
\end{eqnarray}
The physical charged gauge bosons are obtained by diagonalizing this mixing  matrix. However such a matrix is a complex one but we can turn it real through the redefinition, 
\begin{eqnarray}
\hat{V}^+_\mu=e^{i\theta_\chi}V^+_\mu.
\label{Vredefinition}
\end{eqnarray}
Note that this redefinition leaves Eq.~(\ref{Vquarks}) with an explicit phase,
\begin{eqnarray}
{\cal L}^{cc}_{\hat{V}}=-\frac{g}{\sqrt{2}}e^{-i\theta_\chi}\left( \bar{U}_{\beta L}V^{u\dagger}_{4 \beta L}V^d_{3 \gamma L}\gamma^\mu D_{\gamma L} + \bar{U}_{\beta L}V^{uT}_{i \beta L}V^d_{p \gamma L}\gamma^\mu D_{\gamma L}\right) \hat{V}^+_\mu.
\label{newVquarks}	
\end{eqnarray}
As a consequence the interactions among charged gauge bosons and the quarks also violate CP.

We show below  the interactions among quarks and the non-hermitian gauge boson $U^0$, 
\begin{eqnarray}
{\cal L}^{neutral}_U=-\frac{g}{\sqrt{2}}\left( \bar{U}_{\beta L}V^{u\dagger}_{3 \beta L}V^u_{4 \alpha L}\gamma^\mu U_{\alpha L} - \bar{D}_{\gamma L}V^{d\dagger}_{p \gamma L}V^d_{i \sigma L}\gamma^\mu D_{\sigma L}\right) U^0_\mu + \mbox{H.c}.
\label{Uquarks}	
\end{eqnarray}
Again, here the VEV's $v_\chi$  and $v_{\eta^{\prime}}$ promote a mixing among the neutral massive gauge bosons $Z^0$, $Z^{\prime 0}$  and $U^0$  such that the physical bosons come from the diagonalization of such a mixing mass matrix.  As the mixing is due to the VEV's that carry the phases, $v_\chi$  and $v_{\eta^{\prime}}$, the mixing matrix is a complex one. Similarly to the charged gauge boson sector, we can make it real through the redefinition~\cite{coment}
\begin{eqnarray}
	\hat{U}^0=e^{-i\theta_\chi}U^0.
	\label{Uredefinition}
\end{eqnarray}
As an immediate consequence, the interactions  in Eq.~(\ref{Uquarks}) gain a phase,
\begin{eqnarray}
{\cal L}^{neutral}_{\hat{U}}=-\frac{g}{\sqrt{2}}e^{i\theta_\chi}\left( \bar{U}_{\beta L}V^{u\dagger}_{3 \beta L}V^u_{4 \alpha L}\gamma^\mu U_{\alpha L} - \bar{D}_{\gamma L}V^{d\dagger}_{p \gamma L}V^d_{i \sigma L}\gamma^\mu D_{\sigma L}\right) \hat{U}^0_\mu + \mbox{H.c}.
\label{newUquarks}	
\end{eqnarray}

The interactions among quarks and the neutral self-conjugated gauge bosons, $Z^0$  and $Z^{\prime 0}$, always involve quarks and their conjugate field, thus no phase can be present on these interactions.

In summary, in addition to the standard source of CP violation for the quarks, the model disposes  of many other sources of CP violation.  
This implies that several new contributions to the physics of CP violation emerges from this model. However, as can be seen from the results above, all these new CP violating interactions carry at least two new unknown mixing matrix elements, and any process of interest will show up as a large product of these unknown parameters.
In view of this we will stick with just one such effect for illustration purposes. Namely, we will evaluate the contribution to the neutron electric dipole momet (EDM) due to the charged scalars. In FIG.~\ref{fey2} we draw some of the diagrams which enter in the full computation (diagramas with a photon attached to the internal quarks are not shown).
According to the interactions in Eq.~(\ref{CPchargedscalarquarks}), we have contributions to the EDM of a quark already at one loop level. Any such contribution has the general expression~\cite{edm},
\begin{eqnarray}
	d^e_q= \sum_{\beta\,,\,\alpha\,,\,k}\frac{m_\beta}{16\pi^2 m^2_{h_k}}Im( B_{\beta \alpha k}A^*_{\beta \alpha k})\left( Q_\beta J(\frac{m^2_\beta}{m^2_{h_k}})+Q_kI(\frac{m^2_\beta}{m^2_{h_k}}) \right),
	\label{edmgeral}
\end{eqnarray}
where 
\begin{eqnarray}
	I(r)=\frac{1}{2(1-r)^2}\left( 1+r+\frac{2r\ln(r)}{1-r}\right)\,\,\,,\,\,\,J(r)=\frac{1}{2(1-r)^2}\left( 3-r+\frac{2\ln(r)}{1-r}\right),
	\label{factoredm}
\end{eqnarray}
$m_\beta$  and $m_{h_k}$ are the masses of the quarks and scalars, respectively, that run inside the loop and $Q_\beta$ and $Q_k$ are their corresponding electric charges.

We need to consider standard model quarks  $q_{SM}$, whose main contribution to their EDM will be those involving new heavy quarks $T\,,B\,,S$ and the top-quark in the loop. For simplicity, we consider the new heavy quarks only, whose mass scale can be naturally assumed to be around $500$~GeV. As for the scalars we assume their masses of order of $10^2$~GeV. With these choices we have the following prediction for the EDM of $q_{SM}$,
\begin{eqnarray}
d^e_{q_{SM}} \approx 10^{-18}	Im\left( B_{\beta \alpha k}A^*_{\beta \alpha k}\right)e.cm.
\label{SMQEDMprediction}
\end{eqnarray}
We should stress that  $Im\left( B_{\beta \alpha k}A^*_{\beta \alpha k}\right)$ involves the product of small (or at most a fraction of unit) parameters,  six elements of mixing matrix and  two Yukawa couplings. It would be very reasonable to expect that $ Im\left( B_{\beta \alpha k}A^*_{\beta \alpha k}\right)$ could be naturally of the order of $10^{-9}$, which would result in the prediction of $d^e_{q_{SM}} \approx 10^{-27}ecm$.

The valence quark contribution to the neutron EDM is simply obtained from the EDM of the quarks $u$  and $d$ through the relation,
\begin{eqnarray}
	d_n=\frac{4}{3}d_d-\frac{1}{3}d_u.
\label{edmneutron1}	
\end{eqnarray}
For the values of the EDM of the quarks discussed above, we get a prediction for the neutron EDM close to the recent bounds~\cite{edmneutron}, $d_n \approx 10^{-26}e.cm$. Of course, we could have obtained a much smaller value, depending on the smallness of our mixing parameters and Yukawa couplings, but it is interesting that our conservative assumption is fairly able to explain the experimental observation in the context of SCPV in this 3-3-1 model.

\section{lepton sector}

Let us start by the gauge boson interactions among leptons. Forgetting for a while the mixing among the gauge bosons, the redefinitions in Eqs.~(\ref{Vredefinition})  and (\ref{Uredefinition}) lead to the interactions,
\begin{eqnarray}
{\cal L}^{l}_{W,V,U}=-\frac{g}{\sqrt{2}}\left[ \bar \nu_{aL} \gamma^\mu e_{aL} W^+_\mu +e^{-i\theta_\chi}\bar{\nu^C}_{aL} \gamma^\mu e_{aL}\hat{V}^+_\mu +e^{-i\theta_\chi}\bar{\nu^C}_{aL} \gamma^\mu \nu_{aL}\hat{U}^{0 \dagger}_\mu    \right]	+\mbox{H.c}.
\label{leptongaugeboson1}
\end{eqnarray}
These phases in the gauge boson interactions can be easily removed by the redefinition
\begin{eqnarray}
	\bar{\hat{\nu}^C}_{aL}=e^{-i\theta_\chi}\bar{\nu^C}_{aL},
	\label{neutrinoredefinition}
\end{eqnarray}
which gives
\begin{eqnarray}
{\cal L}^{l}_{W,V,U}=-\frac{g}{\sqrt{2}}\left[ \bar \nu_{aL} \gamma^\mu e_{aL} W^+_\mu +\bar{\hat{\nu}^C}_{aL} \gamma^\mu e_{aL}\hat{V}^+_\mu +\bar{\hat{\nu}^C}_{aL} \gamma^\mu \nu_{aL}\hat{U}^{0 \dagger}_\mu    \right]	+\mbox{H.c}.
\label{leptongaugeboson2}
\end{eqnarray}
Thus, differently from  the quark sector, we conclude that the interactions among leptons and gauge bosons do not violate CP.

We have to develop now the Yukawa interactions for the leptons. Opening the last term of  Eq.~(\ref{yukawainteractions}), we obtain
\begin{eqnarray}
	{\cal L}^l_Y=G_{aa}\left[ \bar e_{aL}e_{aR}\rho^0 +\bar \nu_{aL}e_{aR}\rho^+ +\bar{\nu^C}_{aL}e_{aR}\rho^{\prime +
	}  \right]+\mbox{H.c}.
	\label{leptonyukawa}
\end{eqnarray}

Taking the redefinition in Eqs.~(\ref{redchargedscalar}) and (\ref{neutrinoredefinition}), and considering the physical charged scalars given by Eq.~(\ref{physicalchargedscalars}), we get
\begin{eqnarray}
{\cal L}^l_Y=G_{aa}\left[ \bar e_{aL}e_{aR}\rho^0 +e^{-i\theta_\chi}\bar \nu_{aL}{\cal O}^h_{3k}e_{aR}h^+ +e^{i\theta_\chi}\bar{\hat{\nu}^C}_{aL}{\cal O}^h_{4k}e_{aR}h^+ \right]+\mbox{H.c}.
\label{newleptonyukawa}
\end{eqnarray}
Similarly as in the quark sector, CP violation in the leptonic sector arises in the interactions among leptons and the charged scalars of the model. This is a very peculiar result for this kind of model, since it predicts a not yet observed CP violation among leptons. There is a possibility of directly observing these CP violating interactions at next generation of colliders if the charged scalar mixing is not too much suppressed. Of course, their indirect effects, as for instance the dipole moments of leptons, could also be computed by making some assumptions about the couplings and scalar masses, as well as lepton mixing, but this is out of the scope of this study.

\section{conclusions}
We basically implemented the SCPV mechanism in the 3-3-1 model with right-handed neutrinos and pointed the sources of CP violation in the model. A remarkable fact, when compared to other models available, is that this model did not demand any extension for the mechanism to work, neither the potential needed to be the most complete one. In other words, once we departed from a Lagrangian with no complex couplings, thus conserving CP at this level, the existence of complex VEV's inevitably realizes SCPV in the minimal version of the 3-3-1 model with right-handed neutrinos~\cite{2triplets}. It is also important to stress that the model  recover the standard  CKM mixing matrix with three angles and a phase thus recovering all the CP violation effect of the standard model. Other CP violation effects are  genuine effects of new physics at high energies as for example an   EDM for the neutron  around the experimental value. All these new effects  could be computed, in principle, if we had a better knowledge of the several mixing terms and couplings involved, as well as some assumption over the new scalars masses. Besides, it would be possible to see some direct effects at the next generation of colliders since there are CP violating couplings among leptons and charged scalars. Hence, it is a nice outcome to have a neat realization of SCPV in the 3-3-1 model with right handed neutrinos in its minimal version, capable of mimicking the CKM CP violation effects and possessing additional sources of CP violation with consequences beyond the hadronic physics. 

%%%%%%%%%%%%%%%%%%%%%%%%%%%%%%%%%%%%%%%%%%%%%%%%%%%%%%%%%%%%%%%%%%%%%%%%%%%%%%%%%%%%%
\acknowledgments
This work was supported by Conselho Nacional de Desenvolvimento Cient\'{\i}fico e Tecnol\'ogico - CNPq(CASP,PSRS),  Funda\c{c}\~ao de Amparo \`a Pesquisa do Estado 
de S\~ao Paulo - FAPESP(AD).

%%%%%%%%%%%%%%%%%%%%%%%%%%%%%%%%%%%%%%%%%%%%%%%%%%%%%%%%%%%%%%%%%%%%%%%%%%%%%%%%%%%%%%%%%%%
\appendix
\section{}
The mass matrix for the neutral scalar in the basis  $ ( R_\chi \,\, , \,\,
R_{\eta^{\prime}}\,\, , \,\, R_{\chi^{\prime}}\,\, , \,\, R_\eta
\,\, , \,\, R_\rho) $,  is given by
\begin{eqnarray}\begin{tiny}\frac{1}{4}\left(\begin{array}{ccccc} -
\lambda_7\frac{ v_{\eta^{\prime}} v_{\chi^{\prime}} v_\eta }{v_\chi }+f\frac{v_\rho v_{\chi^{\prime}}}{v_\chi} & \lambda_7v_{\chi^{\prime}}v_\eta &
\lambda_7v_\eta v_{\eta^{\prime}} & \lambda_7(v_{\eta^{\prime}}v_{\chi^{\prime}}+2v_\chi v_\eta) &
-fv_{\eta^{\prime}} \\
 - & -\lambda_7\frac{ v_{\eta}v_\chi v_{\chi^{\prime}} v_\eta }{v_{\eta^{\prime}} }+f\frac{v_\rho v_{\chi}}{v_{\eta^{\prime}}} & \lambda_7(v_\chi v_\eta +2v_{\eta^{\prime}}v_{\chi^{\prime}}) & \lambda_7v_\chi v_{\chi^{\prime}} & -fv_{\chi^{\prime}} \\
 - &-
& -\lambda_7\frac{ v_{\eta^{\prime}} v_\chi v_\eta }{v_{\chi^{\prime}}  }-f\frac{v_\rho v_\eta}{v_{\chi^{\prime}} }  & \lambda_7v_\chi v_{\eta^{\prime}}+fv_\rho 
&fv_\eta
  \\
- &-  & -     & -\lambda_7\frac{ v_{\eta^{\prime}}v_\chi v_{\chi^{\prime}}  }{v_{\eta} }-f\frac{v_\rho v_{\chi^{\prime}}}{v_\eta}  & fv_{\chi^{\prime}}\\
- &- & -& -&f\left( \frac{v_{\eta^{\prime}}v_\chi}{v_\rho}-\frac{v_\eta v_{\chi^{\prime}}}{v_\rho}  \right).
\end{array}
\right). \label{matrixRII} \end{tiny}\end{eqnarray}
It is a real mass matrix, thus the mass eigenstates are obtained from the symmetrical eigenstate by an orthogonal transformation
\begin{eqnarray}
\left (
\begin{array}{c}
R_{\chi^{\prime}} \\
R_\eta \\
R_\rho \\
R_\chi \\
R_{\eta^{\prime}}\end{array}
\right )=
{\cal O}^R\left (
\begin{array}{c}
H_1\\
H_2 \\
H_3 \\
H_4 \\
H_5
\end{array}
\right )\,.
\label{HR}
\end{eqnarray}

For the case of the pseudo scalars, we have, in the basis $( I_{\eta^{\prime}} \,\, , I_\chi\,\, , \,\,
I_{\chi^{\prime}}\,\, , \,\, I_\eta \,\, , \,\, I_\rho)$, the following mass matrix
\begin{eqnarray}\begin{tiny} \frac{1}{4}\left(\begin{array}{ccccc} 
 -\lambda_7\frac{ v_{\eta^{\prime}}v_\eta v_{\chi^{\prime}}  }{v_{\chi} }+f\frac{v_\rho v_{\eta^{\prime}}}{v_\chi}& -\lambda_7v_\eta v_{\chi^{\prime}} +fv_\rho & \lambda_7v_\eta v_{\eta^{\prime}} & \lambda_7v_{\chi^{\prime}}v_{\eta^{\prime}} & fv_{\eta^{\prime}} \\
- & -
\lambda_7\frac{v_{\chi^{\prime}}v_\chi v_\eta}{v_{\eta^{\prime}}}+f\frac{v_\chi v_\rho}{v_{\eta^{\prime}}} & \lambda_7v_\eta
v_\chi & \lambda_7
v_{\chi^{\prime}}v_\chi & fv_\chi \\
- & -  & 
-\lambda_7\frac{ v_\eta v_\chi v_{\eta^{\prime}}}{v_{\chi^{\prime}}}-f\frac{v_\eta v_\rho}{v_{\chi^{\prime}}} & -\lambda_7v_{\eta^{\prime}} v_\chi -fv_\rho & -fv_\eta \\
- & - &- & -\lambda_7
 \frac{v_{\chi^{\prime}}v_{\eta^{\prime}}v_\chi}{v_\eta}-f\frac{v_\rho v_{\chi^{\prime}}}{v_\eta} & -fv_\chi \\
- & -
 & - &- &-f\frac{\left( v_\eta v_{\chi^{\prime}}-v_{\eta^{\prime}}v_\chi \right)}{v_\rho}
\end{array}
\right). \label{matrixIII} \end{tiny}\end{eqnarray}
It is also a real mass matrix, thus the mass eigenstates are obtained by an orthogonal transformation
\begin{eqnarray}
\left (
\begin{array}{c}
I_{\chi^{\prime}} \\
I_\eta \\
I_\rho \\
I_\chi \\
I_{\eta^{\prime}}\end{array}
\right )=
{\cal O}^I\left (
\begin{array}{c}
G_1\\
G_2 \\
G_3 \\
G_4 \\
G_5
\end{array}
\right )\,.
\label{IG}
\end{eqnarray}

For the case of charged scalars, we have the following mass matrix  in the basis  $\left( \eta^+ \,,\, \rho^+ \,,\, \chi^+ \,,\,\rho^{\prime +} \right)$,
\begin{eqnarray}\begin{tiny}\frac{1}{2} \left(\begin{array}{cccc} 
\lambda_7( v^2_\chi  +\frac{v_\chi v_{\chi^{\prime}}v_{\eta^{\prime}}}{v_\eta} )-f\frac{v_\rho v_{\chi^{\prime}}}{v_\eta}+\lambda_9v^2_\rho & \lambda_9v_\rho v_\eta -fv_{\chi^{\prime}} & e^{-i\theta_\chi}\lambda_7(v_\eta v_\chi +v_{\chi^{\prime}}v_{\eta^{\prime}}) & e^{-i\theta_\chi}\left( \lambda_9v_\rho v_{\eta^{\prime}}+fv_\chi \right) \\
- & \lambda_8v^2_\chi +\lambda_9v^2_\eta -f(\frac{v_\eta v_{\chi^{\prime}}-v_{\eta^{\prime}}v_\chi}{v_\rho}) & e^{-i\theta_\chi}(v_\chi v_\rho +fv_{\eta^{\prime}}) & e^{-i\theta_\chi}(v_\chi v_{\chi^{\prime}}+v_\eta v_{\eta^{\prime}}) \\
- & - & \lambda_8v^2_\rho -\lambda_7( v^2_{\eta^{\prime}}+\frac{v_{\eta^{\prime}}v_\chi v_\eta}{v_{\chi^{\prime}}}  )-f\frac{v_\eta v_\rho}{v_{\chi^{\prime}}} & \lambda_8v_\rho v_\chi -fv_\eta \\
- & - & - & \lambda_8v^2_{\chi^{\prime}}+\lambda_9v_{\eta^{\prime}}-f\frac{v_\eta v_{\chi^{\prime}}-v_{\eta^{\prime}}v_\chi}{v_\rho}
\end{array}
\right). \label{massmatrix charged} \end{tiny}\end{eqnarray}

This is notoriously a complex mass matrix, but we can easily turn it real by the redefinition of the fields,
\begin{eqnarray}
	\eta^+=e^{i\theta_\chi}\hat{\eta}^+\,\,\,\,\,\, \mbox{and}\,\,\,\,\,\,\rho^+=e^{i\theta_\chi}\hat{\rho}^+.
	\label{redchargedscalar}
\end{eqnarray}
In this case an orthogonal transformation takes the symmetric eigenstates into the mass eigenstate
\begin{eqnarray}
\left (
\begin{array}{c}
\chi^+ \\
\hat{\eta}^+ \\
\hat{\rho}^+ \\
\rho^{\prime +}\end{array}
\right )=
{\cal O}^h\left (
\begin{array}{c}
h^+_1\\
h^+_2 \\
h^+_3 \\
h^+_4
\end{array}
\right )\,.
\label{physicalchargedscalars}
\end{eqnarray}
\begin{figure}[!htb]
\centerline{\hbox{
   \epsfxsize=0.9\textwidth
   \epsfbox{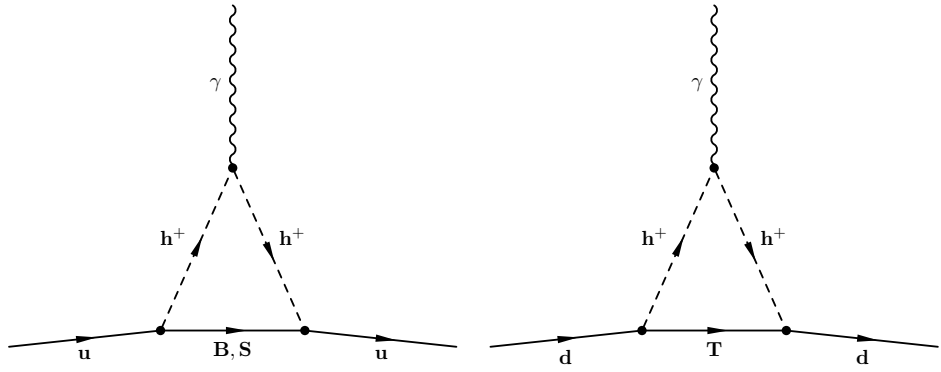} 
     }
  } 
      \caption{ Some of the one-loop diagrams contributing to neutron EDM.}
\label{fey2} 
\end{figure}


\begin{thebibliography}{99}
\bibitem{coment1}
A similar mechanism may arise in the leptonic sector, in an extension of SM including massive neutrinos,  if the element $U_{e3}$ of the neutrino mixing matrix is not zero.
\bibitem{coment}
This kind of mixing promoted by $v_\chi$ or $v_{\eta^{\prime}}$  in the neutral sector of the gauge bosons is found in the papers of Ref. \cite{2triplets}.
\end{thebibliography}
\end{document}